\documentclass{aastex631}

\usepackage{bm}
\usepackage{comment}
\usepackage{xcolor}
\received{\today}
\submitjournal{ApJ}

\shorttitle{Effects of Alfv\'enic fluctuations on Magnetic Ejecta}
\shortauthors{Scolini et al.}
\graphicspath{{./}{Figures/}}
\begin{document}

\title{Can Alfv\'enic Fluctuations Affect the Correlation and Complexity of Magnetic Fields in Magnetic Ejecta? \\
A Case Study Based on Multi-Spacecraft Measurements at 1~au}

\correspondingauthor{Bin Zhuang}
\email{bin.zhuang@unh.edu}

\author[0000-0002-5681-0526]{Camilla Scolini}
\altaffiliation{Current affiliation: European Research Council Executive Agency, Brussels, Belgium}
\affiliation{Institute for the Study of Earth, Oceans, and Space, University of New Hampshire, Durham, NH, USA}

\author[0000-0002-5996-0693]{Bin Zhuang}
\affiliation{Institute for the Study of Earth, Oceans, and Space, University of New Hampshire, Durham, NH, USA}

\author[0000-0002-1890-6156]{No\'e Lugaz}
\affiliation{Institute for the Study of Earth, Oceans, and Space, University of New Hampshire, Durham, NH, USA}

\author[0000-0002-9276-9487]{R\'eka M. Winslow}
\affiliation{Institute for the Study of Earth, Oceans, and Space, University of New Hampshire, Durham, NH, USA}

\author[0000-0001-8780-0673]{Charles J. Farrugia}
\affiliation{Institute for the Study of Earth, Oceans, and Space, University of New Hampshire, Durham, NH, USA}

\author[0000-0001-5731-8173]{Norbert Magyar}
\affiliation{Centre for mathematical Plasma Astrophysics, KU Leuven, Leuven, Belgium}

\author[0000-0002-7526-8154]{Fabio Bacchini}
\affiliation{Centre for mathematical Plasma Astrophysics, KU Leuven, Leuven, Belgium}
\affiliation{Royal Belgian Institute for Space Aeronomy, Brussels, Belgium}

%%%%%%%%%%%%%%%%%%%%%%%%%%%%%%%%%%%%%%%%%%%%%%%%%%%%%%%%%%%%
\begin{abstract}
We investigate whether Alfv\'enic fluctuations (AFs) can affect the structure of magnetic ejecta (MEs) within interplanetary coronal mass ejections (ICMEs). We study an ICME observed on 2001 December 29 at 1 au by ACE and Wind, at a total angular separation of $\sim$0.8$^\circ$ ($\sim0.014$~au). We focus on the correlation and complexity of its magnetic structure measured between two spacecraft in association with large-amplitude AFs. The Alfv\'enicity of the ME is investigated in terms of the residual energy and cross helicity of fluctuations. We find that as for the event of interest, large-amplitude AFs occur in the rear region of the ME at both Wind and ACE with a duration of about six hours. We compare the correlation of the magnetic field strength and vector components measured between Wind and ACE, and investigate complexity in terms of the magnetic hodograms. The region showing AFs is found to be associated with a decreased correlation of the magnetic field components and an increased complexity of the ME magnetic configuration detected at ACE and Wind, which may be due to the fact that the two spacecraft crossing the same ME along different trajectories likely sampled AFs in different oscillation phases. Combining multi-point in-situ measurements and remote-sensing observations of the ICME source region, we further discuss different potential sources of the AFs.

%find that the AFs detected near the ME rear likely originated in the interplanetary space as a result of interaction with a following large-scale solar wind transient.
\end{abstract}



%%%%%%%%%%%%%%%%%%%%%%%%%%%%%%%%%%%%%%%%%%%%%%%%%%%%%%%%%%%%
\section{Introduction} 
\label{sec:introduction}
%%%%%%%%%%%%%%%%%%%%%%%%%%%%%%%%%%%%%%%%%%%%%%%%%%%%%%%%%%%%

% - ICME coherence and correlation
The question of whether interplanetary coronal mass ejections \citep[ICMEs;][]{Bothmer1998, Cane2003, Kilpua2017}, especially their magnetically dominated substructure, referred as magnetic ejecta \citep[ME;][]{Kilpua2017} can behave as coherent structures, is a long-standing problem \citep[e.g.][]{Burlaga1981, Owens2017}. Coherence corresponds to the fact that the structure is capable of responding to external perturbations in a collective manner, which has broad implications for the global evolution of ICME magnetic structures, interpretation of single-spacecraft in-situ measurements, and ICME space weather predictions based on instruments located upstream of Earth \citep{laker2024,regnault2024}. 

% - the role of AFs as mediators of anti-coherence based on Paper I
The very definition of magnetic coherence \citep[][]{Owens2017} implies that a coherent behavior can only be exhibited if information about the acting perturbation is able to timely propagate across an ICME structure. Since ICMEs are low plasma-beta structures,  Alfv\'en waves become a very likely mediator or carrier that transfers the perturbation information \citep{Owens2017}.
In a recent paper, \citet{Scolini2023b} found that the role of low-frequency (between $\sim$~$10^{-5}$ and $10^{-3}$~Hz), large-amplitude Alfv\'enic fluctuations (AFs), which can present extended periods covering $\sim$20\% of the ICME regions, can be a possible mediator of the ICME coherent behavior based on a set of 10 ICMEs measured at 1~au by multiple spacecraft in the vicinity of L1.  They also found that the occurrence of AFs are associated with a decreased correlation level in the magnetic field components of the ICME measured at different spacecraft, in which the magnetic field correlation of a single CME measured at multiple spacecraft was used to investigate whether the CME magnetic field structure can evolve coherently in past studies \citep[e.g.,][]{Matsui2002,Lugaz2018,Ala-Lahti2020,Farrugia2023}.

% - the next step - this paper
The study of \citet{Scolini2023b} was aimed at drawing an initial picture of the role of AFs as possible mediators of ICME magnetic field correlation based on the largest possible number of ICMEs in longitudinal scales of 0.008--0.011~au (or $0.5^\circ$--$0.7^\circ$). The consideration of such scale range enables us to investigate how AFs alter the ICME internal substructure on mesoscales based on multi-spacecraft in-situ measurements with available separations so far (at an angular separation $<1^\circ$ but not too small to obscure differences of magnetic field correlation, e.g., $>0.5^\circ$).
Despite our intention of being as inclusive as possible, the nature of the topic investigated required us to impose restrictive criteria on the selection of events. This resulted in a small selection of 10 ICMEs detected at 1~au as homogeneous as possible in terms of in-situ characteristics (e.g., the presence of a preceding shock and sheath, the existence of a well-defined magnetic flux rope structure) and in terms of the relative position of the observing spacecraft. 
Furthermore, the causality of the detected anti-correlation between Alfv\'enicity and magnetic field correlation in the preceding study requires further investigation, i.e., contextualization of single-point in-situ measurements with respect to the solar wind conditions in which the ICME propagates and of the evolution of the ICME. Therefore, the close physical relationship between AFs and the correlation of magnetic field signatures through the detailed investigation of an ICME case are required.

% - advantages of case studies 
The advantage of considering a single event in detail lies in the possibility of exploiting the variety of observational signatures that characterize the Sun--to--Earth propagation of a given ICME as well as locating the origin of the AFs and understanding their impact on the ICME structure. In addition to investigating multi-spacecraft measurements of the plasma and magnetic field parameters, it is useful to compare the direction of propagation of AFs with the magnetic connectivity between the ICME and Sun using suprathermal electron signatures.
The connection between AFs and the magnetic connectivity of ICMEs measured at 1~au was previously investigated by \citet{Borovsky2019} and \citet{Good2020, Good2022}. They discussed how the Alfv\'enic properties of ICMEs and solar wind at 1~au may be partly determined based on the degree to which the magnetic field lines in coronal source regions are open or closed.

% paper structure
In this work, we focus on an ICME observed on 2001 December 29 and study the close physical relationship between AFs and the correlation as well as the complexity of the magnetic field components of the ICME. The paper is structured as follows. 
Section~\ref{sec:data_and_methods} provides an overview of the datasets and methods used to quantify the Alfv\'enicity and correlation between the ICME properties at different spacecraft locations. 
Section~\ref{sec:results_event1} presents event observations and analyses of its in-situ signatures in the context of understanding the relationship between magnetic field correlation and complexity in association with AFs. 
In Section~\ref{sec:discussion}, we consider the characteristics of the source region to construct a global picture of the ICME structure and explore different scenarios for the origin of the observed AFs.
In Section~\ref{sec:conclusions}, we summarize our findings and draw our conclusions. 

%%%%%%%%%%%%%%%%%%%%%%%%%%%%%%%%%%%%%%%%%%%%%%%%%%%%%%%%%%%%
\section{In-Situ Data and methods}
\label{sec:data_and_methods}
%%%%%%%%%%%%%%%%%%%%%%%%%%%%%%%%%%%%%%%%%%%%%%%%%%%%%%%%%%%%

%%%%%%%%%%%%%%%%%%%%%%%%%%%%%%%%%%%%%%%%%%%%%%%%%%%%%%%%%%%%
\subsection{Spacecraft Positions and Data}
\label{subsec:data}

\begin{figure}[ht!]
\centering
\includegraphics[width=0.95\linewidth]{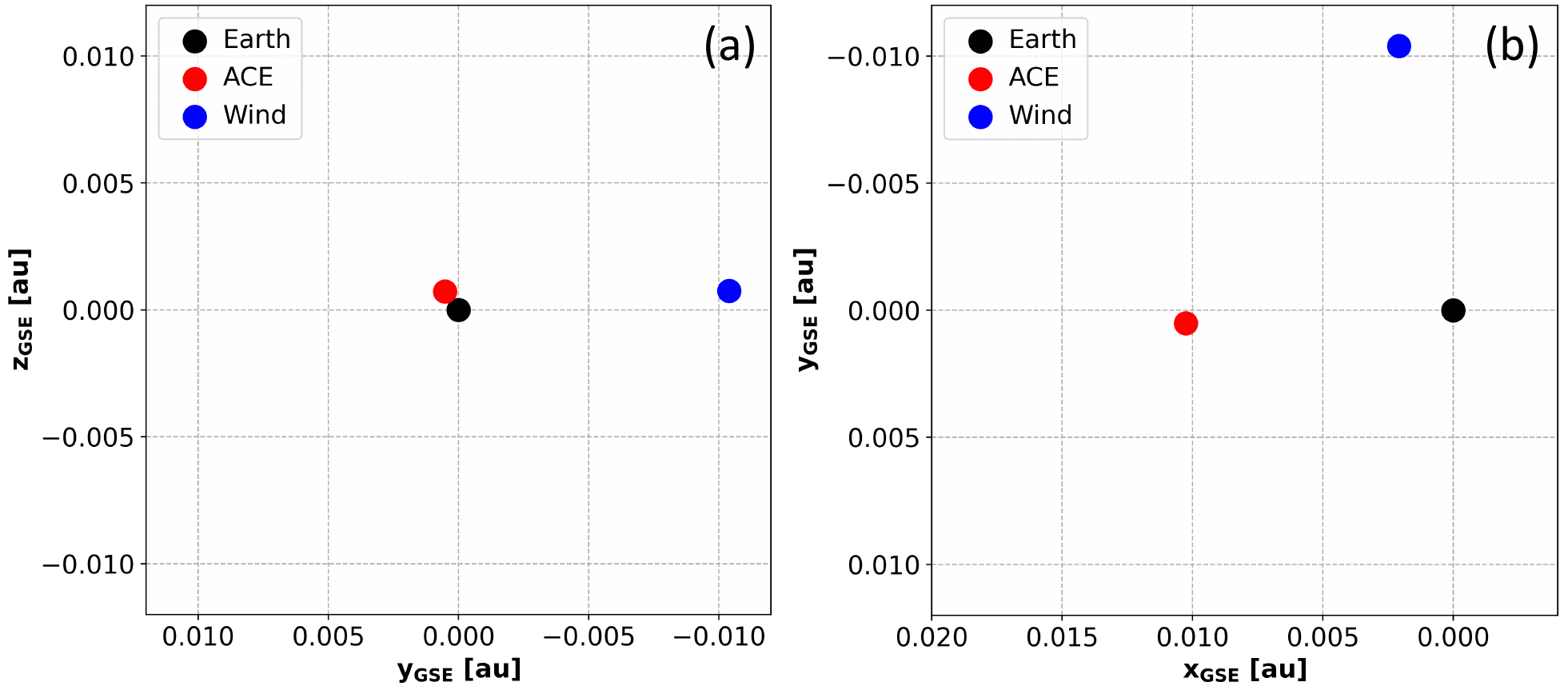} % positions on December 29, 2001 at 05:00 UT. 
\caption{Relevant spacecraft positions in GSE coordinates at 05:00 UT on 2001 December 29.
(a): view in the Y$_\mathrm{GSE}$-Z$_\mathrm{GSE}$ plane, looking towards the Sun.
(b): view in the X$_\mathrm{GSE}$-Z$_\mathrm{GSE}$ plane, looking downwards on the ecliptic plane.
Earth, ACE, and Wind are shown in black, red, and blue, respectively. Large separation in the east-west direction (Y) corresponds to Wind's
second distant prograde orbit.}
\label{fig:event1_spacecraft_positions}
\end{figure}

% spacecraft position
We investigate the ICME arriving at ACE and Wind on 2001 December 29 % ICME start and end times
that is listed as event 7 in Table~1 of \citet{Scolini2023b}.
At that time, ACE was located at ($240.60$, $12.04$, $17.11$)~Earth radii ($R_E$) in Geocentric Solar Ecliptic (GSE) coordinates.
Wind was on its second distant prograde orbit where it ventured a long way from Earth in the Y$_\mathrm{GSE}$-direction, and was located at ($48.815$, $-243.81$, $17.59$)~$R_E$. This corresponds to a total separation of $\sim 320$~$R_E$ ($\sim 0.014$~au or $\sim 0.80^\circ$), and a Y-separation of $\sim 256$~$R_E$ ($\sim 0.011$~au or $\sim 0.63^\circ$).\
Figure~\ref{fig:event1_spacecraft_positions} illustrates the spacecraft positions relative to Earth on 2001 December 29 at 05:00 UT. 

% dataset used
In the analysis, we use ACE magnetic field data at 16-s cadence taken by the Magnetic Field Experiment \citep[MAG;][]{Smith1998},
and measurements of the solar wind plasma and suprathermal ($\sim$272~eV) electron pitch-angle distribution (PAD) at 64-s cadence from the Solar Wind Electron Proton Alpha Monitor \citep[SWEPAM;][]{McComas1998}. 
At Wind, we use measurements of the magnetic field at 3-s cadence taken by the Magnetic Field Investigation \citep[MFI;][]{Lepping1995}, 
complemented by measurements of the solar wind plasma properties at 3-s cadence, and of suprathermal ($\sim$265~eV) electron PAD data at 24-s cadence from the Three-Dimensional Plasma and Energetic Particle Investigation \citep[3DP;][]{Lin1995}.

\subsection{Identification of Alfv\'enic Fluctuations} %%%%%%%%%%%%%%%%%%%%%%%%%%
\label{subsec:methods_af_identification}

We analyze the ICME Alfv\'enicity using an approach similar to that of \citet{Scolini2023b}, which is summarized below.
To identify periods of AFs from in-situ measurements, %in and around the ICME (including MEs, preceding sheaths, and surrounding solar wind), 
the velocity and magnetic field fluctuations are explored through wavelet analysis of the normalized residual energy:
\begin{equation}
    \sigma_r(k,t)=\frac{E_v(k,t)–E_b(k,t)}{E_v(k,t)+E_b(k,t)},
\end{equation}
where $E_v(k,t)$ and $E_b(k,t)$ are the sum of the power of the wavelet transforms \citep{Torrence1998} of the vector components of the velocity $\bm{v}(t)$ and magnetic field $\bm{b}(t)$ 
 given in velocity units of $\bm{b}(t)=\bm{B}(t)/\sqrt{\mu_0 \rho}$ (with $\rho$ denoting the particle density), respectively \citep[][]{Telloni2012, Telloni2013, Good2020, Telloni2021, Good2022}, and are functions of time $t$ and wavenumber $k$. 
$\sigma_r(k,t)$ measures the relative importance between the kinetic and magnetic energies, and is expected to be close to zero in a reference frame co-moving with the solar wind due to the equipartition of the magnetic and kinetic energies of potentially existing AFs. 
This method enables us to investigate long periods of data through visual inspection across wide frequency ranges and is employed to identify a set of candidate AF periods. 

Information on the predominant direction of propagation of candidate AF periods with respect to the background magnetic field direction is further derived from the normalized cross helicity:
\begin{equation}
    \sigma_c(k,t)=\frac{E_{+}(k,t)–E_{–}(k,t)}{E_{+}(k,t)+E_{–}(k,t)},
\end{equation}
where $E_\pm(k,t)$ is the sum of the power of the wavelet transforms of the components of the Els\"asser variables $\bm{z_\pm(t)} = \bm{v}(t) \pm \bm{b}(t)$. The condition of $\sigma_c(k,t)<0$ ($>0$) indicates a propagation predominantly parallel (anti-parallel) to the local magnetic field direction, and $\sigma_c(k,t)\simeq 0$ indicates a balanced propagation along both directions. 

We perform the wavelet analysis using the Paul wavelet \citep[due to its better time localization capability compared to the Morlet wavelet;][]{Telloni2012} and considering a period of two days before the ICME start and two days after the ME end in order to avoid effects related to the cone of influence (the region of the wavelet spectrum where edge effects become important) at the edges of the time period of interest. 
We perform the analysis on both ACE and Wind data. Before applying the wavelet transforms to the magnetic field and plasma time-series data, we resample them to a common cadence at both ACE and Wind. The resampling cadence is dictated by the lowest cadence available across all data sets at both Wind and ACE, i.e.\ 64~s according to ACE/SWEPAM.

On the basis of the wavelet spectra, at each time step we integrate $\sigma_r(k,t)$ and $\sigma_c(k,t)$ by computing their median values across scales $k_i$ corresponding to periods between 30~min and 12~hours \citep[approximately between $5.5 \times 10^{-4}$ and $2.3\times 10^{-5}$~Hz, and falling within the injection range of the power spectrum; see][]{Good2020}. In this way, we obtain time-dependent medians for $\sigma_r(k,t)$ and $\sigma_c(k,t)$ which are purely functions of time $t$. 

\subsection{Correlation of Multi-Point Magnetic Field Measurements} %%%%%%%%%%%%%%%%%%%%%%%%%%
\label{subsec:methods_correlation}

Similarly to \citet{Scolini2023b}, we compute the correlation between the time profiles of the magnetic field strength and components within the MEs measured at Wind and ACE. To do so, for each event we first take the shock time and ME boundaries at Wind as reference. 
The sheath and ME time-series portions at ACE are then each shifted and stretched to match the sheath and ME start and end times at Wind. During this process, the Wind magnetic field data is first rebinned to 16-s averages to match with the ACE data. We analyze the correlation in two ways.

% global correlation
First, we compute the global correlation of the magnetic field strength and magnetic field components within the ICME between Wind and ACE. The calculation of the global correlation considers all data points in a specific period, e.g., the sheath or ME periods. In just the global correlation calculation case, in order to a) ensure enough data points for the correlation calculation and b) neglect disturbances in high-resolution data, both Wind and ACE data sets are rebinned to 5 minutes. The correlation is computed for the magnetic field strength and the three magnetic field components and is provided in the form of global Pearson correlation coefficients $\vec{cc} = (cc_B, cc_{B_R}, cc_{B_T}, cc_{B_N})$. This approach enables us to reduce the relation between the two groups of signals to a single value.

% time-dependent correlation
Second, to gain insight into how the magnetic field correlation is distributed throughout different ICME sub-structures, we further explore the instantaneous (i.e.\ time-dependent) Pearson correlation between ACE and Wind time series as a function of different temporal scales. We do so in both the sheath and ME regions individually by measuring the Pearson correlation between Wind and ACE starting from a small portion of the signal, and then repeating the process along a rolling window until the entire structure is covered. When using rolling windows, we must exclude a portion of data that corresponds to the window size. We perform the computation from the start to the end and vice versa, and use an average to represent the related coefficient if there exist a pair of forward and backward calculated coefficients at a certain time step. The window size $\Delta t_i$ is set to equally increase from 30~minutes (to be consistent with the smallest temporal scale explored in studying AFs in Section~\ref{subsec:methods_af_identification}) to 6~hours (considering the duration of the region associated with AFs inside the ICME of interest) with an increment of 5~minutes. The medians of $(cc_B, cc_{B_R}, cc_{B_T}, cc_{B_N})$ within the periods between 30 minutes and 6 hours are also calculated.

\subsection{Combining $\sigma_c$ and Suprathermal Electron PAD} %%%%%%%%%%%%%%%%%%%%%%%%%%
\label{subsec:methods_pads}

When complemented with suprathermal electron PAD data giving information on the ME global magnetic field topology, $\sigma_c$ can also serve to constrain the origin of AFs \citep[e.g.][]{Good2020}. In the presence of counter-streaming electron strahls, indicating a closed ICME magnetic field topology, AF periods with $\sigma_c\sim0$ could suggest a solar origin of AFs, which would have been generated before the CME passes the Alfv\'en surface. In contrast, unbalanced $\sigma_c$ periods suggest an interplanetary origin. In the latter case, the spatial origin of AFs with respect to an observer can be further inferred from the direction of propagation of the dominant wave mode along the ME magnetic field. 
Alternatively, the presence of uni-directional suprathermal electrons (indicative of an ICME magnetic topology being disconnected from the Sun at one leg) may be used to discriminate solar and interplanetary AF origins based on their propagation direction relative to the ME axial magnetic field and AFs. In addition, we consider $\sigma_c$ in the interpretation of multi-point AF observations to, e.g., determine if AFs observed at multiple locations may have the same origin.

%%%%%%%%%%%%%%%%%%%%%%%%%%%%%%%%%%%%%%%%%%%%%%%%%%%%%%%%%%%%
\section{The ICME on 2001 December 29}
\label{sec:results_event1}

\subsection{Overview of In-situ Measurements} %%%%%%%%%%%%%%%%%%%%%%%%%%
\label{subsec:results_overview}

% in situ signatures
The event under consideration is an ICME observed at Wind and ACE starting from 2001 December 29. 
This ICME is included in the Richardson and Cane ICME list \citep[hereafter RC ICME catalog;][\url{https://izw1.caltech.edu/ACE/ASC/DATA/level3/icmetable2.html}]{Richardson2010}, in the Wind ICME catalog \citep[][\url{https://wind.nasa.gov/ICME_catalog/ICME_catalog_viewer.php}]{NievesChinchilla2018}, as well as in the ACE ICME catalog \citep[][\url{https://izw1.caltech.edu/ACE/ASC/DATA/level3/ICME_List_1995_2009_Jian.pdf}]{Jian2006a}.
% solar counterpart 
The RC ICME catalog mentions a possible association of the ICME under study with the CME erupted on December 26 at 05:30 UT from \citep[NOAA AR 9742, located around N10W60 on the solar disk; see also][]{Grechnev2016}. This CME is listed in the CDAW CME catalog (\url{https://cdaw.gsfc.nasa.gov/CME_list/}) as a partial halo CME with a linear speed of 1446 km~s$^{-1}$. Despite the source region being $60^\circ$ to the west, this was a fast and therefore likely wide CME that had the potential to impact Earth. This is also the most probable solar counterpart to the ICME under consideration based on a careful investigation of all other CMEs observed within one week prior to the ICME arrival at 1~au. 

\begin{figure}[ht!]
\centering
\includegraphics[width=\linewidth]{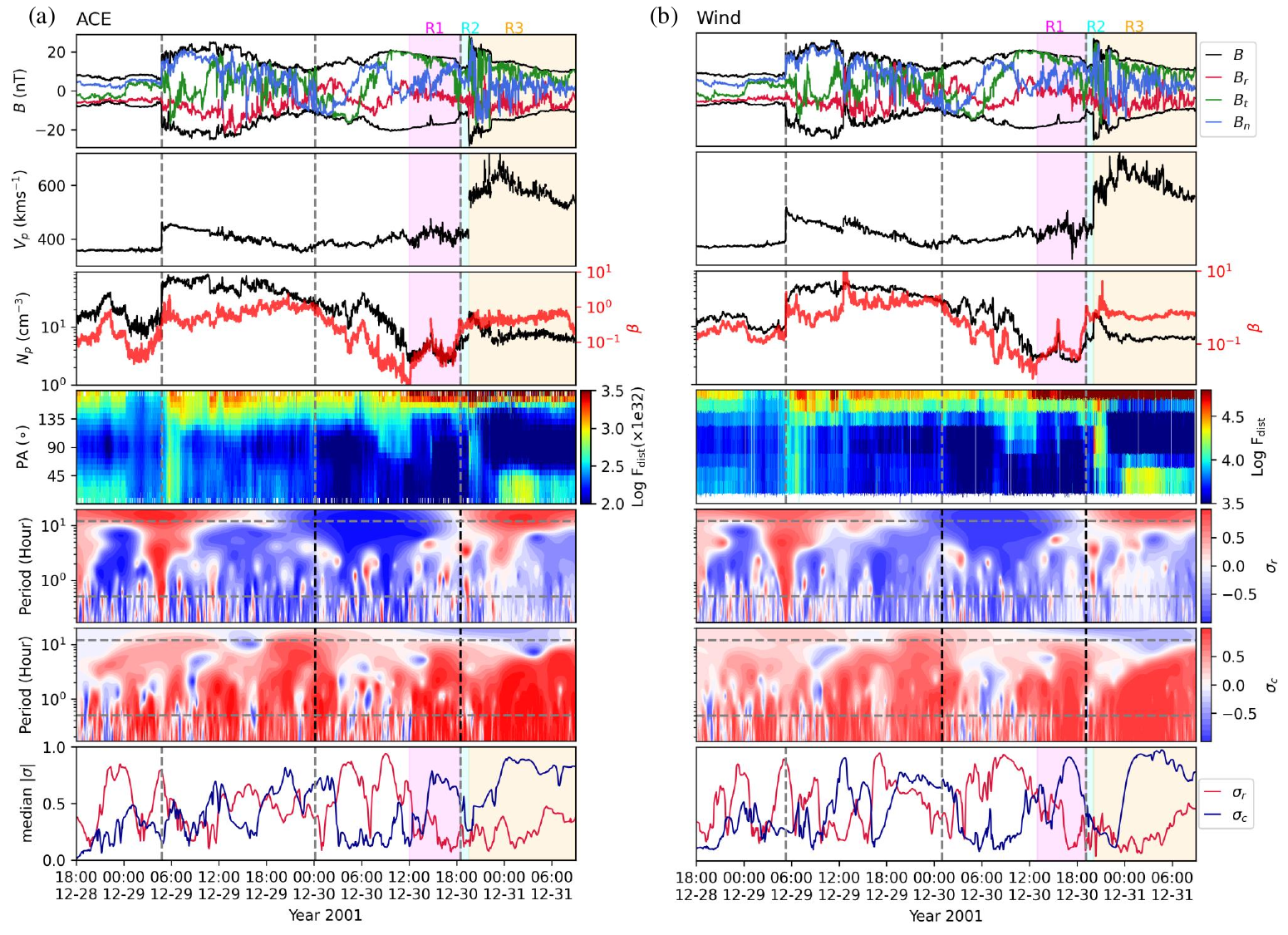}
\caption{In situ plasma and magnetic field signatures for the 2001 December 29 ICME.
(a): ACE. (b): Wind.
From top to bottom: magnetic field components, radial speed, proton number density associated with proton beta (in red), suprathermal electron PAD data, wavelet of $\sigma_r(k, t)$, wavelet of $\sigma_c(k, t)$, and median of $|\sigma_r(k, t)|$ and $|\sigma_c(k, t)|$ across scales $k_i$ between 30~minutes and 12~hours. Velocity and magnetic field data are provided in RTN coordinates. The vertical dashed lines mark the shock time and ME boundaries at each spacecraft. Highly Alfv\'enic regions R1, R2, and R3 are marked by the shaded magenta, cyan, and orange areas, respectively. The two horizontal dashed lines in the $\sigma_r$ and $\sigma_c$ plots indicate the injection range between 30 minutes and 12 hours.}
\label{fig:event1_insitu}
\end{figure}

% in situ properties
The ICME signatures at 1~au in radial-tangential-normal (RTN) coordinates are shown in Figure~\ref{fig:event1_insitu} at ACE (a) and Wind (b). In this case, we use Wind/3DP data instead of Wind/SWE for the plasma measurements because there exist Wind/SWE data gaps near the ICME end \citep[Figure~1 of][]{Scolini2023b}.
At ACE, ICME signatures started with the passage of an interplanetary shock on December 29 at 04:47 UT, followed by an ME with flux rope signatures starting on December 30 at around 00:10 UT. The ME ended on December 30 around 18:30 UT, indicated by the decrease in the total magnetic field strength. With this choice of boundaries at ACE, the sheath had a duration of 19.38~hours, and the ME had a duration of 18.33~hours.
At Wind, the interplanetary shock was detected on December 29 at 05:17~UT, while the ME started on December 30 at around 01:00~UT and ended at around 19:09~UT on the same day.
With this choice of boundaries at Wind, the sheath had a duration of 19.72~hours, and the ME had a duration of 18.15~hours.

\begin{figure}[ht!]
 \centering
 \includegraphics[width=\linewidth]{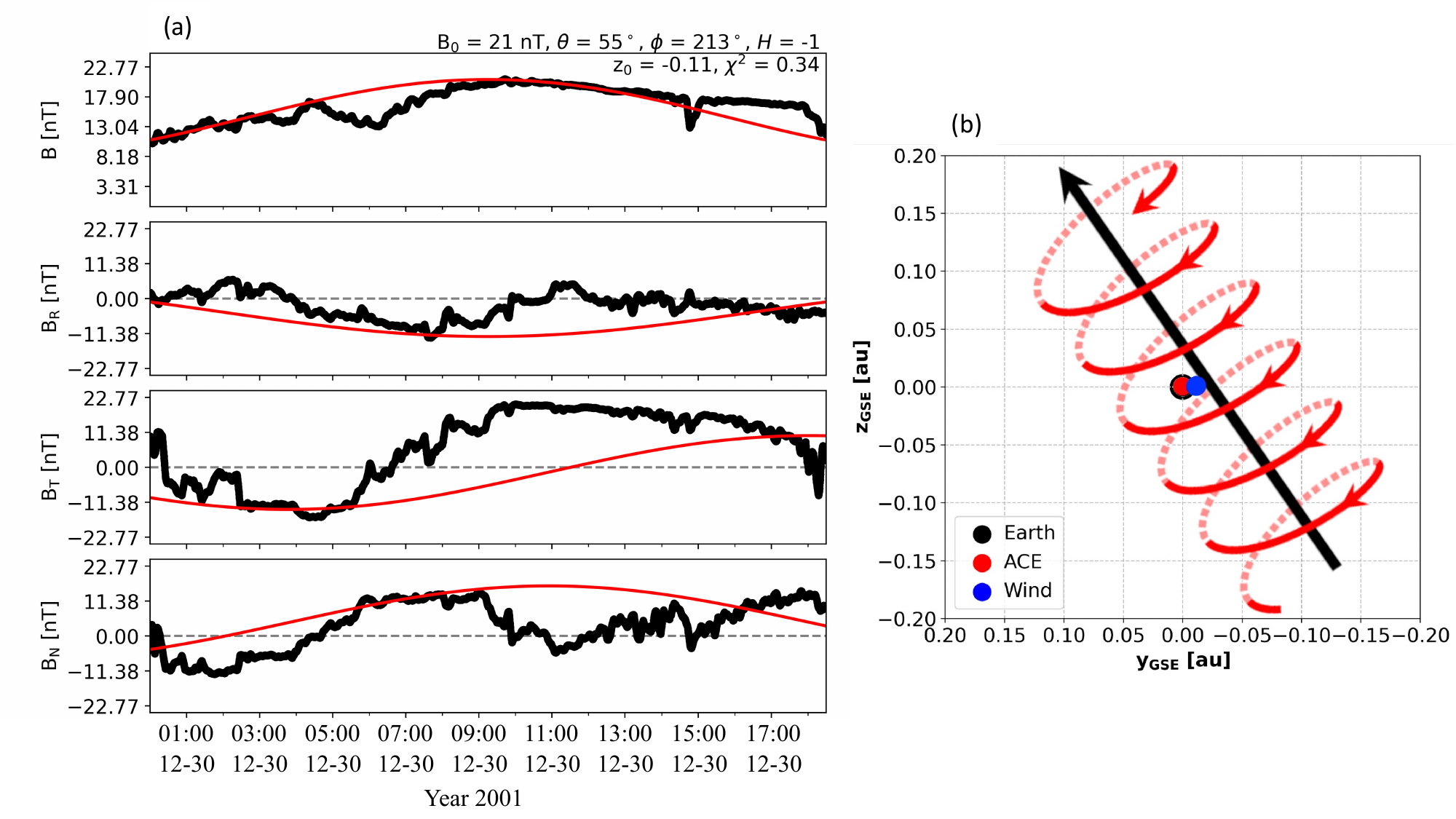}
\caption{
Modeling of the ICME flux rope magnetic field configuration based on the in-situ magnetic field data at ACE.
(a): LFF fit.
(b): crossing of Earth (black), Wind (blue) and ACE (red) through the ME flux rope magnetic structure as reconstructed using a LFF fit.
}
\label{fig:event1_lff_fit}
\end{figure}

The ME at both ACE and Wind presents clear rotations in the magnetic field components (the first row). % consistent with an $F+$ FR type.
From visual inspection, $B_N$ rotates from negative to positive values (i.e.\ from south to north), while $B_T$ rotates from negative to positive values (i.e.\ from east to west). If assuming a simple, twisted flux rope configuration, these rotations suggest that the flux rope has a negative helicity and is moderately inclined between a SEN and an ENW flux rope type \citep[e.g.][]{Bothmer1998, Kilpua2017}. We further perform a fit of the magnetic configuration using the linear force-free (LFF) flux rope model \citep{Lepping1990} at both Wind and ACE. 
The fitted results are shown in Figure~\ref{fig:event1_lff_fit}~(a) for ACE data. The LFF fit models the flux rope as having a left-handed chirality (related to negative helicity) and an axial field direction of $(\theta, \phi)_\mathrm{LFF}^\mathrm{ACE} = (55^\circ, 213^\circ)$ (all fitted angles are given in RTN coordinates). The spacecraft impact parameter (normalization of the closest-approach distance with the flux rope radius) is estimated to be -0.11. Similar results (not shown) are obtained for Wind data, where the axial direction is estimated to be 
$(\theta, \phi)_\mathrm{LFF}^\mathrm{Wind} = (40^\circ, 220^\circ)$. 
Figure~\ref{fig:event1_lff_fit}~(b) represents the flux rope geometry to scale with respect to the spacecraft positions at 1~au. We note that the LFF model can not fully capture of the flux rope structure of the ME, particularly in relation to the $B_T$ component and the complex  magnetic fields. Complex magnetic configurations in different ME sub-regions, especially in the ME rear portion corresponding to the appearance of AFs are further investigated using magnetic hodograms as presented in Section~\ref{subsec:results_complexity}.

% preceding ICME/long sheath
Based on past studies \citep[e.g.,][]{Lepping1995, Richardson2010}, the ME at 1~au typically lasts about 1~day and has a radial size of 0.2--0.3~au, while the ME sheath typically lasts about 12~hours and has a radial size shorter than 0.1~au. Compared with these typical values, the ICME in this study presents a very long sheath region (19.38 and 19.72~hours at ACE and Wind, respectively). The sheath radial size (estimated from its duration and average speed) is 0.2~au, which is significantly longer than the typical sheath size. The ME, on the other hand, has a radial size of 0.18~au, which is slightly smaller than average but not atypical.
We consider that the long duration of the sheath may be due to a) the interaction with (or cannibalism of) a preceding ICME or b) the flank crossing of the ICME. This first hypothesis is consistent with the fact that the RC ICME catalog identifies a preceding ICME on December 28, although this ICME is not listed in the Wind and ACE ICME catalogs. We leave further investigation of this long-lasting sheath to future works.

% following ICME/fast wind
From Figure~\ref{fig:event1_insitu} we also see that the ME is followed by a faster solar wind propagating at a speed of around 600~km~s$^{-1}$. This faster wind is behind a shock most likely driven by a following ICME, as identified in the Wind ICME catalog (shock arrival on December 30 at 20:05 UT, with ME signatures between on December 30 at 23:10 UT and December 31 at 07:46 UT). The same ICME is also listed in the ACE ICME catalog as associated with a shock detected on December 30 at 19:33 UT, and ME signatures ending on December 31 at 07:47 UT (no ME start time is given). Most likely due to the high plasma beta and lack of flux rope signatures, no following ME is listed in the RC ICME catalog. At both ACE and Wind, this second shock presented a shock normal direction that is almost perfectly aligned with the radial direction ($\hat{n} = [-0.98, -0.16, 0.08]$ at Wind, and $\hat{n} = [-1.00, 0.07, 0.02]$ at ACE, from the IPshocks catalog: \url{https://ipshocks.helsinki.fi}). Such a radial shock normal is typically observed in association with ICME-driven shocks rather than shocks driven by other interplanetary structures \citep[e.g., stream interaction regions; see][]{Huang2019, Ala-Lahti2020}, which is consistent with the picture of this shock being driven by a following ICME. 
% alfvenic nature of this second structure
% Regardless of this shock's precise origin, our focus is on exploring how the highly Alfv\'enic nature of the following fast wind may have contributed to the formation of the AFs in the rear portion of the ME.

% ME magnetic connectivity
Regarding magnetic connectivity of the ME to the Sun, we can observe this in the fourth row in Figure~\ref{fig:event1_insitu} that the ME lacks bi-directional suprathermal electron signatures, with only the presence of the electron strahl at around $180^\circ$. This suggests that only one ICME leg is still connected to the Sun by the time the structure reached 1~au. Based on the LFF modeling shown in Figure~\ref{fig:event1_lff_fit}~(b), this is most likely the easternmost leg, and the westernmost leg is likely to have already disconnected.

% Alfvenicity -- R1
The wavelet spectra and especially the time-series profiles of median $|\sigma_r|$ in Figure~\ref{fig:event1_insitu} indicate the presence of a period of high Alfv\'enicity in the rear part of the ME at both Wind (from 13:27 UT to 19:10 UT on December 30) and ACE (from 12:37 UT to 18:30 UT on December 30). This period, hereafter referred to as R1, is highlighted in magenta in Figure~\ref{fig:event1_insitu}. Based on the time-dependent medians of $\sigma_r$ and $\sigma_c$ in the bottom two rows of the figure, R1 shows $\sigma_r$ values close to 0, especially at ACE, with an average of $-0.09$ at ACE and $-0.29$ at Wind. R1 presents high $\sigma_c$ (average of 0.54 at ACE and 0.61 at Wind), indicating that at both spacecraft AFs propagate predominantly in the direction anti-parallel to the local magnetic field. We note that the duration of the R1 region accounts for $\sim$30\% of the duration of the whole ME.
For comparison, the portion of the ME enclosed between the ME start time and the beginning of R1 presents average $\sigma_r$ values deviated from 0 ($-0.40$ at ACE and $-0.63$ at Wind) and lower than the averages in R1 and the averages as listed in Table~2 in \citet{Scolini2023b}, which indicates a lack of AFs in this region.

% R2
High Alfv\'enicity contents are also present in a narrow region between the end of the ME and the following interplanetary shock (hereafter called R2 and marked in cyan in Figure~\ref{fig:event1_insitu}). This region presents $\sigma_r$ values close to 0 (average of 0.12 at ACE and 0.09 at Wind) together with average $\sigma_c$ values of 0.38 at ACE and 0.29 at Wind, suggesting the presence of a mixture of counter-propagating and uni-directional AFs.
% R3
Finally, AFs are also found to occur in the fast solar wind downstream of the following interplanetary shock (hereafter called R3 and indicated in orange in Figure~\ref{fig:event1_insitu}), which presents $\sigma_r$ values close to 0 (0.10 at ACE and -0.11 at Wind) together with quite high $\sigma_c$ values (average of 0.59 at ACE and 0.54 at Wind), indicating the presence of AFs counter-propagating with respect to the local magnetic field.

\subsection{Magnetic Field Correlation and Alfv\'enicity} %%%%%%%%%%%%%%%%%%%%%%%%%%
\label{subsec:results_correlation}

\begin{figure}[ht!]
	\centering
	\includegraphics[width=0.92\textwidth]{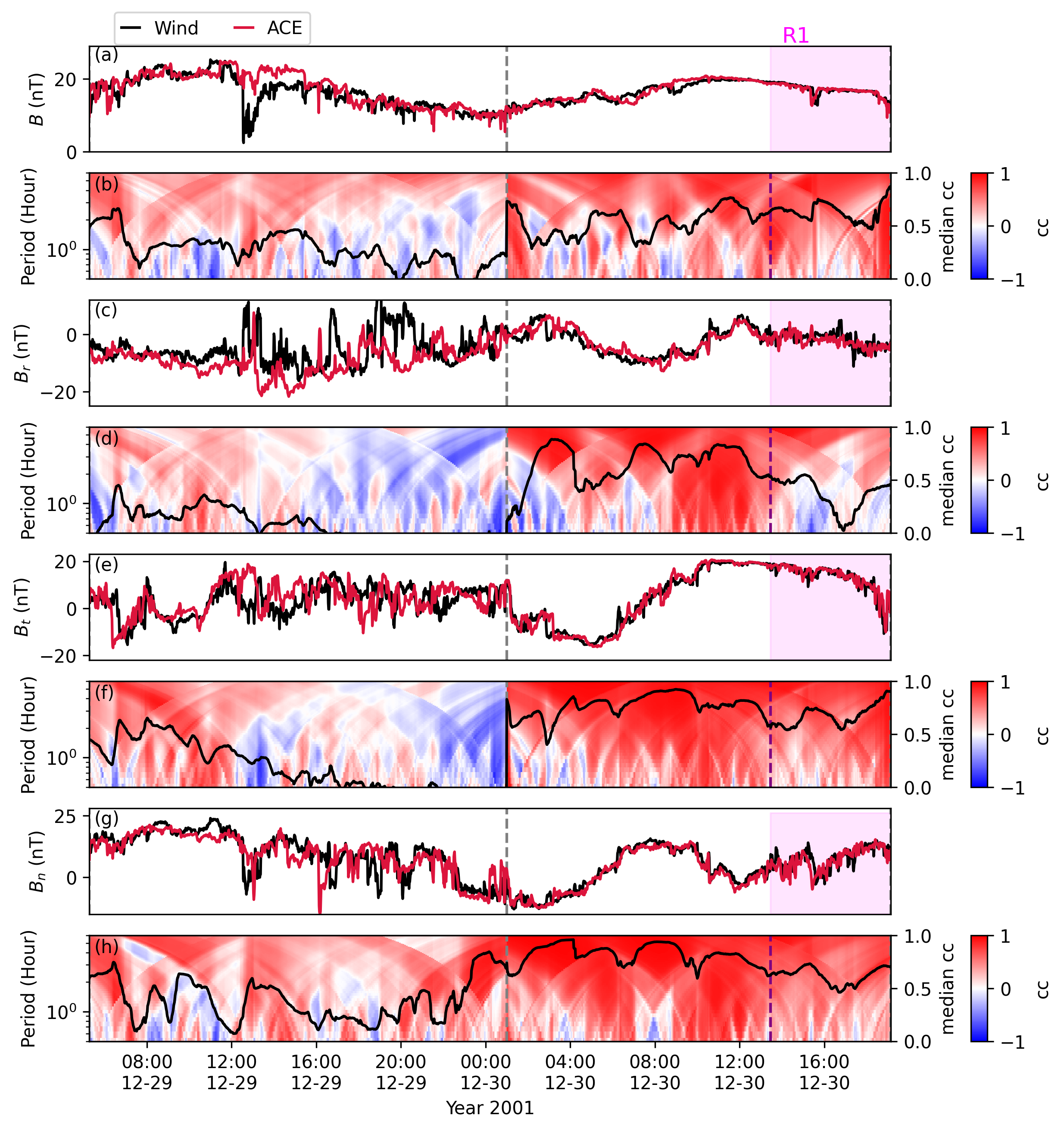}
	\caption{Correlation of magnetic field signatures at ACE and Wind for the 2001 December 29 ICME.
		(a), (c), (e), and (g): $B$, $B_R$, $B_T$, and $B_N$ profiles at Wind (in black) and ACE (in red). 
		ACE profiles are time-shifted to the ME start time at Wind and stretched to match the ME end time at Wind.
		(b), (d), (f), and (h): spectra of instantaneous correlation of magnetic field components at ACE and Wind as a function of the temporal scale considered. The black solid curves in the spectrum plot correspond to the median values of the correlation coefficients within the periods between 30 minutes and 6 hours. The black dashed line marks the ME start at Wind. The region corresponding to R1 is shown as the magenta-shaded area. 
	}
	\label{fig:event1_correlation_multiscale}
\end{figure}

% magnetic field correlation
Next, we investigate whether a relationship exists between the Alfv\'enicity and the correlation of magnetic field profiles detected between ACE and Wind, as shown in Figure~\ref{fig:event1_correlation_multiscale}.

% ME correlations
We start by evaluating the correlation in the magnetic field profiles detected at ACE and Wind in panels~(a), (c), (e), and (g), finding that the ME presents very high average global correlations in magnetic field strength and all components ($cc_{B} = 0.91$, $cc_{B_R} = 0.81$, $cc_{B_T} = 0.97$, $cc_{B_N} = 0.94$). In fact, the ME region preceding R1 presents global correlation coefficients of $cc_{B} = 0.92$, $cc_{B_R} = 0.85$, $cc_{B_T} = 0.98$, and $cc_{B_N} = 0.97$, which are similar to those calculated across the whole ME. Conversely, R1 presents global correlation coefficients that are lower than those for the average ME for all magnetic field components: $cc_{B}=0.82$, $cc_{B_R} = 0.48$, $cc_{B_T} = 0.90$, and $cc_{B_N} = 0.68$.

We then estimate the time-dependent magnetic field correlation based on different time windows ranging from 30~minutes to 6~hours, as also shown in panels~(b), (d), (f), and (h) in Figure~\ref{fig:event1_correlation_multiscale}. The coefficient spectra of the three magnetic field components (especially $B_R$) reveal that the R1 region where large-amplitude AFs are present has lower coefficient levels compared to the region enclosed between the starts of ME and R1. This is further supported by the solid curves corresponding to the median values of the correlation coefficients within the periods between 30 minutes and 6 hours in the spectrum plots. In R1, the averages of the coefficients between 30 minutes and 6 hours are for 0.33, 0.61, and 0.56 for $B_R$, $B_T$, and $B_N$, respectively, compared to 0.57, 0.69, and 0.73 for the region enclosed between the beginnings of ME and R1.

% sheath correlations
Figure~\ref{fig:event1_correlation_multiscale} also indicates that the sheath exhibits lower global correlations with $cc_{B} = 0.74$, $cc_{B_R} = 0.19$, $cc_{B_T} = 0.38$, $cc_{B_N} = 0.79$. These lower correlations are consistent across all temporal scales considered, which is consistent with the general picture given in \citet{Scolini2023b} through the superposition of magnetic field correlation profiles from 10 ICMEs at 1~au.

\begin{figure}[ht!]
	\centering
	\includegraphics[width=0.92\linewidth]{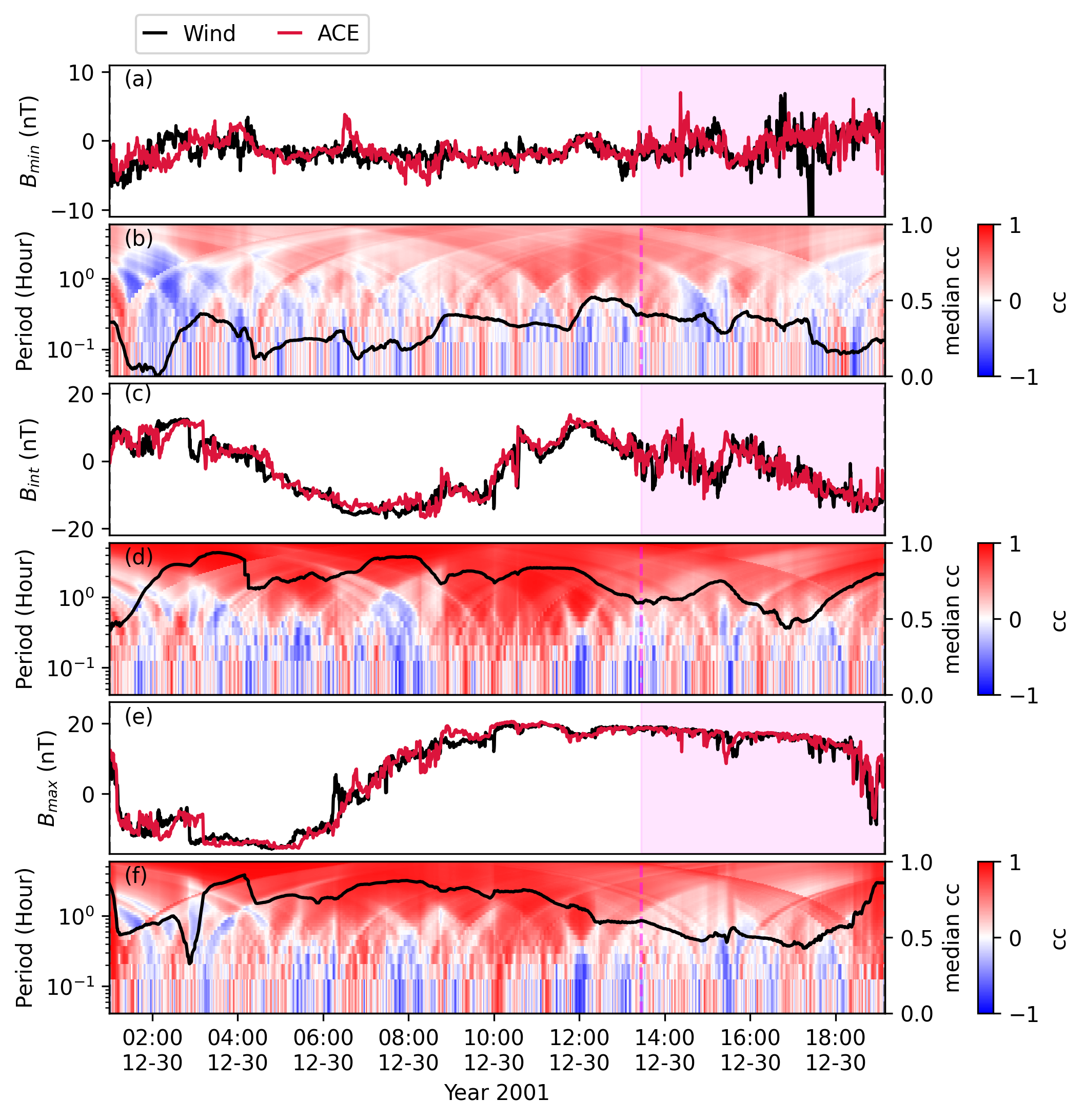}
	\caption{Correlation of magnetic field signatures at ACE and Wind for the 2001 December 29 ICME.
		(a), (c), (e): $B_{min}$, $B_{int}$, and $B_{max}$ profiles at Wind (in black) and ACE (in red). 
		ACE profiles are time shifted to the ME start time at Wind and stretched to match the ME end time at Wind.
		(b), (d), (f): instantaneous correlation of magnetic field components at ACE and Wind as a function of the temporal scale considered. The black solid curves in (b), (d), and (f) correspond to the median values of the correlation coefficients within the periods between 30 minutes and 6 hours. The R1 region is shown as the magenta shaded area. 
	}
	\label{fig:event1_correlation_multiscale_mva}
\end{figure}

% projection to MVA frame for the ME
We further perform a minimum variance analysis \citep[MVA;][]{Sonnerup1998} on the ME structure and re-evaluate the correlation between magnetic field signatures at Wind and ACE after having projected the magnetic field signatures in the MVA frame at each spacecraft. In the MVA frame, approximately corresponding to the frame of the flux rope, the magnetic field components are projected in the $min$, $int$, and $max$ directions corresponding to the directions of minimum, intermediate, and maximum variance. For a flux rope structure with the spacecraft's crossing path close to the axis, the $int$ direction corresponds to the direction of its magnetic axis, $max$ corresponds to the poloidal direction of the magnetic field vector, and $min$ completes the right-handed triad. 
We find MVA axis directions of $(\theta,\phi)_\mathrm{MVA}^\mathrm{Wind} = (55^\circ, 211^\circ)$ at Wind, and $(\theta,\phi)_\mathrm{MVA}^\mathrm{ACE} = (55^\circ, 214^\circ)$ at ACE, which are consistent with the LFF modeling at both spacecraft. 

The correlation in the magnetic field components after projecting to the MVA frame is shown in Figure~\ref{fig:event1_correlation_multiscale_mva}. We obtain global correlations of 
$cc_{B_{min}} = 0.41$, $cc_{B_{int}} = 0.92$, and $cc_{B_{max}} = 0.98$ for the whole ME region, in which 
correlation of $B_{min}$ is smaller compared to the other two components. 
When looking at the region before the start of R1, we find correlations similar to those of the whole ME region: 
$cc_{B_{min}} = 0.38$, $cc_{B_{int}} = 0.95$, and $cc_{B_{max}} = 0.98$.
However, within R1, we find lower correlations: 
$cc_{B_{min}} = 0.27$, $cc_{B_{int}} = 0.79$, and $cc_{B_{max}} = 0.87$. Similarly, the spectra of the component correlation show a lower correlation level in the R1 region, associated with decline in the median of the coefficient as shown by the solid curves in the spectrum plots. The average coefficients for $B_{min}$, $B_{int}$, and $B_{max}$ in R1 are 0.28, 0.60, and 0.51, respectively. While the correlation coefficient of $B_{min}$ is 0.26, the  coefficients of $B_{int}$, and $B_{max}$ in R1 are smaller than those of in the region between the ME start and R1 start (0.72 and 0.69, respectively).

Overall, the correlations in the pre-R1 and R1 regions within the ME clearly hint towards the existence of an anti-correlation between the Alfv\'enicity (evaluated in terms of $\sigma_r$) and the correlation of the magnetic field component profiles detected at ACE and Wind. This is consistent with the interpretation proposed by \citet{Scolini2023b} that the presence of AFs within a structure may actually lower the spatial correlation of the magnetic field properties at different points due to the fact that the measurement of the magnetic field components at Wind and ACE are associated with the AFs during different oscillation phases. 

\subsection{ME Complexity and Alfv\'enicity} %%%%%%%%%%%%%%%%%%%%%%%%%%
\label{subsec:results_complexity}

Another aspect of MEs that has been explored in recent research efforts is the so-called ``complexity'' \citep[e.g.\ ][and references therein]{Winslow2022}.  
The absolute complexity of an ME at any one heliocentric distance is difficult to define in isolation because it requires a reference state of assumed low complexity. Generally speaking, however, complexity can be understood as the degree of similarity or deviation of a given ME structure from a ``standard'' configuration characterized by a flux-rope magnetic structure connected back to the Sun by two ``legs'' \citep[see, e.g., Figure~2 in][]{Zurbuchen2006}.

\begin{figure}[ht!]
\centering
\includegraphics[width=1\linewidth]{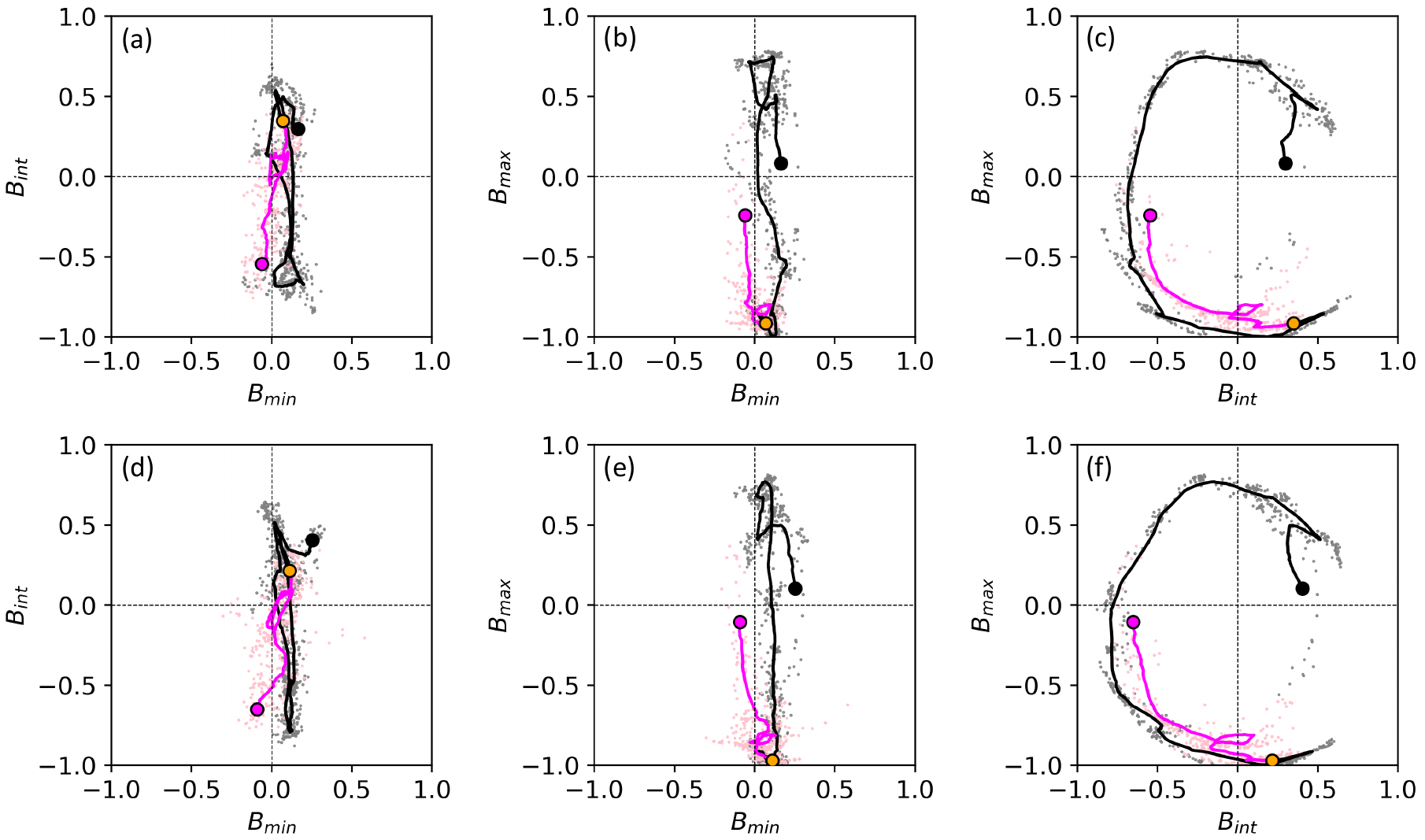}
\caption{Magnetic hodograms of the ME at ACE (top) and Wind (bottom) for the 2001 December 29 ICME. At each spacecraft, the magnetic field components are projected in the respective MVA frame. Signatures for the ME pre-R1 region are shown in black, while those for the R1 region are shown in magenta.
The start of the ME is marked by the black circle, the start of R1 by the orange circle, and the end of the ME/R1 by the magenta circle.}
\label{fig:event1_magnetic_hodograms}
\end{figure}

As discussed in Section~\ref{subsec:results_overview}, from visual inspection the ME in this study appears to present a flux rope signature.
However, a closer look at the structure reveals that rotations in the magnetic field components not only exceed the $180^\circ$ expected from a spacecraft crossing a single twisted flux rope structure, but actually present an inverted sense of rotation in the R1 region. This behavior is clearly visible in the hodograms of the magnetic field components at ACE and Wind after projecting to their respective MVA frame, as shown in Figure~\ref{fig:event1_magnetic_hodograms}. According to the classification scheme proposed by \citet{NievesChinchilla2018}, the ME signatures at both ACE and Wind possess characteristics of a $C_x$ (complex) class, presenting more than one rotation \citep{NievesChinchilla2019}. The temporal correspondence between the $C_x$ signatures and the AFs in R1 is particularly intriguing as it opens new scenarios regarding the possible relation between large-amplitude AFs and the magnetic configuration of MEs. In this event, we observe that the development of a complex magnetic signature temporally coincides with the presence of strong AFs near the back of the ME. %Although the exact cause-effect relationship between these two factors is hard to establish based solely on observations at 1~au, we can see that the inversion in the rotation of the magnetic field components within the R1 region (at both Wind and ACE) is actually simultaneous with the appearance of AFs. %To further clarify whether such an inversion is actually a signature of the AFs populating R1, we take a closer look at the non-radial velocity signatures within the ME and following solar wind at ACE and Wind, as shown in Figure~\ref{fig:event1_velocity}.

% such higher non-radial flows may be due to the interaction, at least shock-ejecta interaction
%
%Moreover, these non-radial flows exhibit a global oscillatory behavior with periodicity of about 3-to-6~hours which is in phase with the large-scale magnetic field fluctuations visible in $B_T$ and $B_N$ in the R1 region in Figure~\ref{fig:event1_insitu}.
%This indicates that the low $\sigma_r$ and high $\sigma_c$ values within R1 are actually a manifestation of a long-period population of AFs within R1, and are not simply due to the variation of the background magnetic field. This, in turn, suggests that the complex magnetic field configuration revealed by the magnetic hodograms in Figure~\ref{fig:event1_magnetic_hodograms} is actually the result of the presence of AFs in R1. In that sense, this indicates that, at least in specific situations such as the one considered here, AFs can be associated with an increased ME complexity as observed in situ by individual spacecraft.

%  VERY NICE !!! - Thanks Charlie!

%%%%%%%%%%%%%%%%%%%%%%%%%%%%%%%%%%%%%%%%%%%%%%%%%%%%%%%%%%%%
\section{Reconstructing the global scenario} 
\label{sec:discussion}
%%%%%%%%%%%%%%%%%%%%%%%%%%%%%%%%%%%%%%%%%%%%%%%%%%%%%%%%%%%%

\subsection{Connecting In-situ Magnetic Signatures to the Source Region Magnetic Configuration} %%%%%%%%%%%%%%%%%%%%%%%%%%
\label{subsec:results_global_magnetic_configuration}

\begin{figure}[ht!]
\centering
\includegraphics[width=\linewidth]{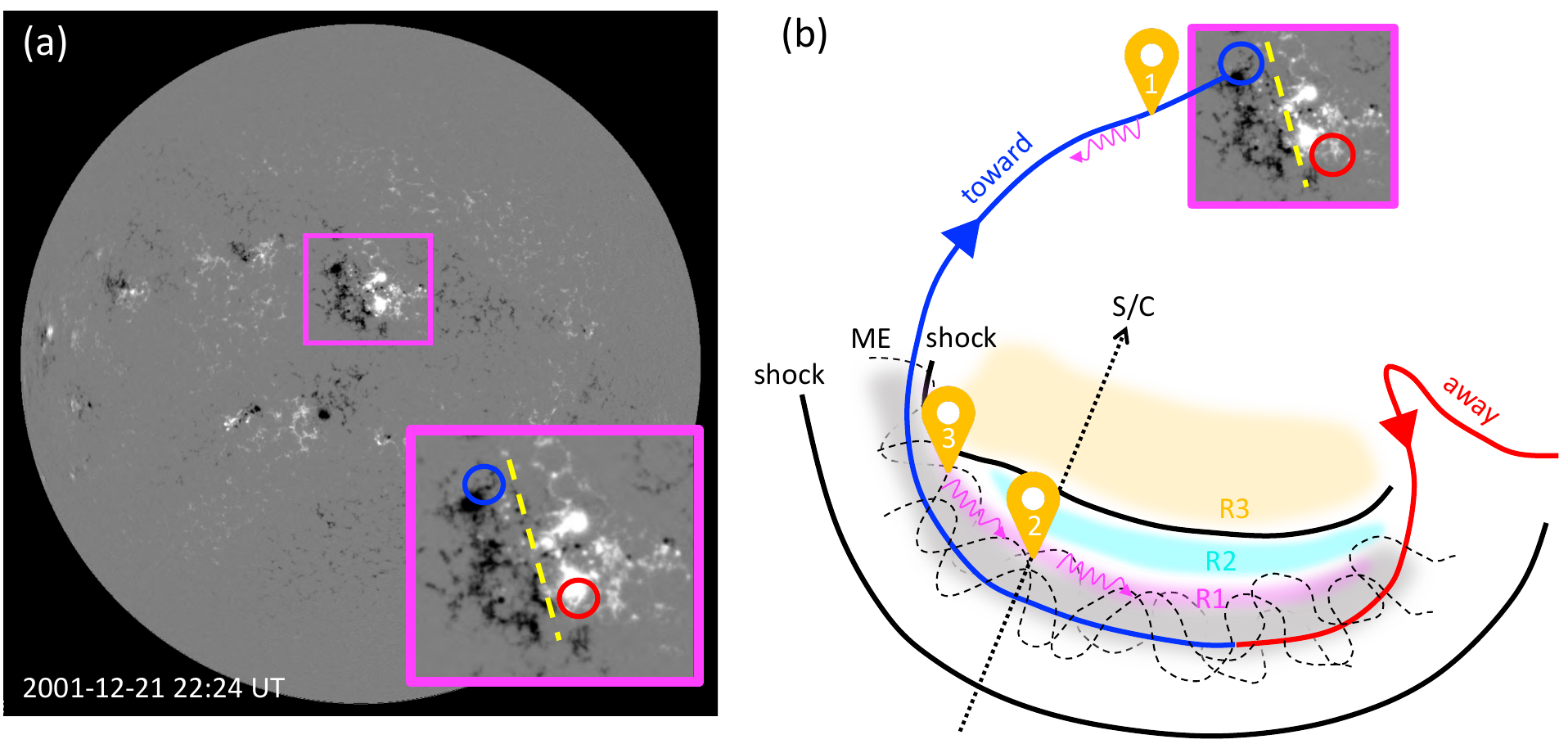}
\caption{Source region magnetic configuration and global magnetic topology for the 2001 December 29 ICME.
(a): SOHO/MDI photospheric magnetic field map on 2001 December 21 at 22:24 UT. AR 9742 is zoomed onto in the magenta box, with the polarity inversion line marked in yellow, and possible CME footpoints marked by the blue (negative polarity) and red (positive polarity) circles. 
(b): sketch representing the global magnetic topology of the ME at 1 au.
Three possible source locations for the AFs in R1 are marked by the three orange pins.}
\label{fig:event1_source_region}
\end{figure}

To place the reconstruction of the ICME magnetic structure based on in-situ information into a global context, we further explore the magnetic field configuration of the active region (AR) from which the event under investigation is likely to have originated. 
Magnetic field observations of AR 9742 from the Solar and Heliospheric Observatory (SOHO) Michelson Doppler Imager (MDI) are shown in Figure~\ref{fig:event1_source_region}~(a). 
The image is taken a few days before the CME eruption, on December 21, when the AR crossed the central meridian. 
We see a clear bipolar magnetic configuration, with a highly inclined polarity inversion line and negative magnetic polarity on the east and a positive magnetic polarity on the west. SOHO Extreme-Ultraviolet Imaging (EIT) telescope in 195~\AA \, (not shown) also shows coronal loops skewed in reverse-J shapes, which suggest that the AR had a negative magnetic helicity, making it likely for the erupted flux rope to also have a negative chirality. The picture is consistent with the flux rope configuration and magnetic polarity inferred from in-situ measurements and the LFF fit of the ME structure, with the inward magnetic field mapping to the easternmost ME leg. Suprathermal electron signatures also suggest that by the time the ICME reached 1~au, its eastern leg was likely the only one to remain anchored to the Sun. A schematic of the reconstructed global magnetic field configuration of the ICME while at 1~au is shown in Figure~\ref{fig:event1_source_region}~(b).

\subsection{Locating the Origin of AFs in R1} %%%%%%%%%%%%%%%%%%%%%%%%%%
\label{subsec:results_alfvenicity}

The direction of the ME axial magnetic field as reconstructed from the LFF fitting in Figure~\ref{fig:event1_lff_fit}~(b) and the positive $\sigma_c$ within R1 suggest that AFs in R1 are globally propagating from the northeast to the southwest direction, opposite to the ME axial field direction. 
Thus, it is reasonable to expect that their origin location must also lie to the east of both Wind and ACE. 
The presence of the suprathermal electron strahls propagating in the same direction as AFs in R1 (opposite to the local magnetic field) suggests that, in principle, these AFs may either have formed in the interplanetary space, or near the Sun at the footpoint of the easternmost ME leg in the early propagation phases (i.e.\ before the CME crossed the Alfv\'en surface). These possible source locations are marked by the orange pins in Figure~\ref{fig:event1_source_region}~(b).

% option 1: solar origin %%%%%%%%%%%%%%%%%%%%

AFs of solar origin could be detected at 1~au, marked by pin number 1 in Figure~\ref{fig:event1_source_region}~(b), only if they originated in the very initial propagation phases before the CME crossed the Alfv\'en critical point and the AFs became ``frozen-in'' and swept out with the supersonic flow at higher altitudes.
Thus, AFs formed at such an early stage are expected to to have enough time to involve a large portion of the eruptive structure and to be observed in interplanetary space as a balanced state with AFs propagating both parallel and anti-parallel to the flux rope axis (regardless of whether either of the footpoints/legs of an ICME has become disconnected from the Sun during propagation through the interplanetary space). %Furthermore, AFs of solar origin would be expected to involve the whole ME structure in the radial direction, rather than just a localized portion of the ME \citep[][]{Good2022}.
Due to the fact that R1 is very localized at the back of the ME and presents highly uni-directional AF signatures suggests an interplanetary origin for the AFs within R1, while a solar origin may be unlikely. 

% interplanetary origin %%%%%%%%%%%%%%%%%%%%
Next, we consider the hypothesis of interplanetary formation for the AFs in R1.
%
% option 2: IP local formation due to velocity shears or magnetic reconnection
One possibility is that AFs within R1 formed locally in situ at both spacecraft (marked by pin number 2) due to interactions with the solar wind immediately following the ME (corresponding to R2). 
As possible AF formation mechanisms in the solar wind, we focus our attention on magnetic reconnection and velocity shears, which have been identified as relevant to the formation of AFs in other space plasma contexts \citep[particularly in the solar corona;][]{Velli1999, Kigure2010, Cranmer2018}. 
By inspecting the magnetic field and plasma signatures within R1 and at the boundary between R1 and R2 using the high-resolution data at Wind, we could find no signatures \citep[see e.g.,][]{gosling2005} of magnetic reconnection occurring locally near L1 that contributed to the generation of AFs in R1. However, it is noted that reconnection between the CME and background magnetic field \citep[also called erosion;][]{Dasso2006,Lavraud2014,Ruffenach2015,Farrugia2023} can occur when the CME is closer to the Sun and then weaken or cease when reaching 1~au. Erosion can peel off the outermost magnetic field lines of the ME and results in an imbalance in the ME azimuthal magnetic flux. Following the method in \citet{Dasso2006} and \citet{Farrugia2023}, we calculate the azimuthal magnetic flux from the front to the end of the ME, and find that the flux is imbalanced (not shown here), which indicating the occurrence of erosion during the CME propagation. In comparison to the symmetry of azimuthal magnetic field component in an ideal condition without erosion, the asymmetry in $B_{max}$ as shown in panel~(e) of Figure~\ref{fig:event1_correlation_multiscale_mva} indicates that the erosion occurs at the ME front. Besides, the sheath region is found to be much longer than the typical sheath size, which also supports the erosion at the ME front and indicates that a portion of the sheath is the ME origin. The narrow region with high Alfv\'enicity ($\sigma_r$ is close to 0 and $\sigma_c$ is positively larger than 0.6) near the ME region may be related to the front erosion. If the erosion also contributed to the AFs in the rear region of R1, it would suggest that the AFs generated near the front propagate along and anti-parallel to the local magnetic field line (e.g., $B_n$) from the front to the end. The opposite signs of $B_n$ and the same positive $\sigma_c$ near the front and end can support the scenario of the front--to--rear propagation of AFs.

\begin{figure}[ht!]
	\centering
	\includegraphics[width=\linewidth]{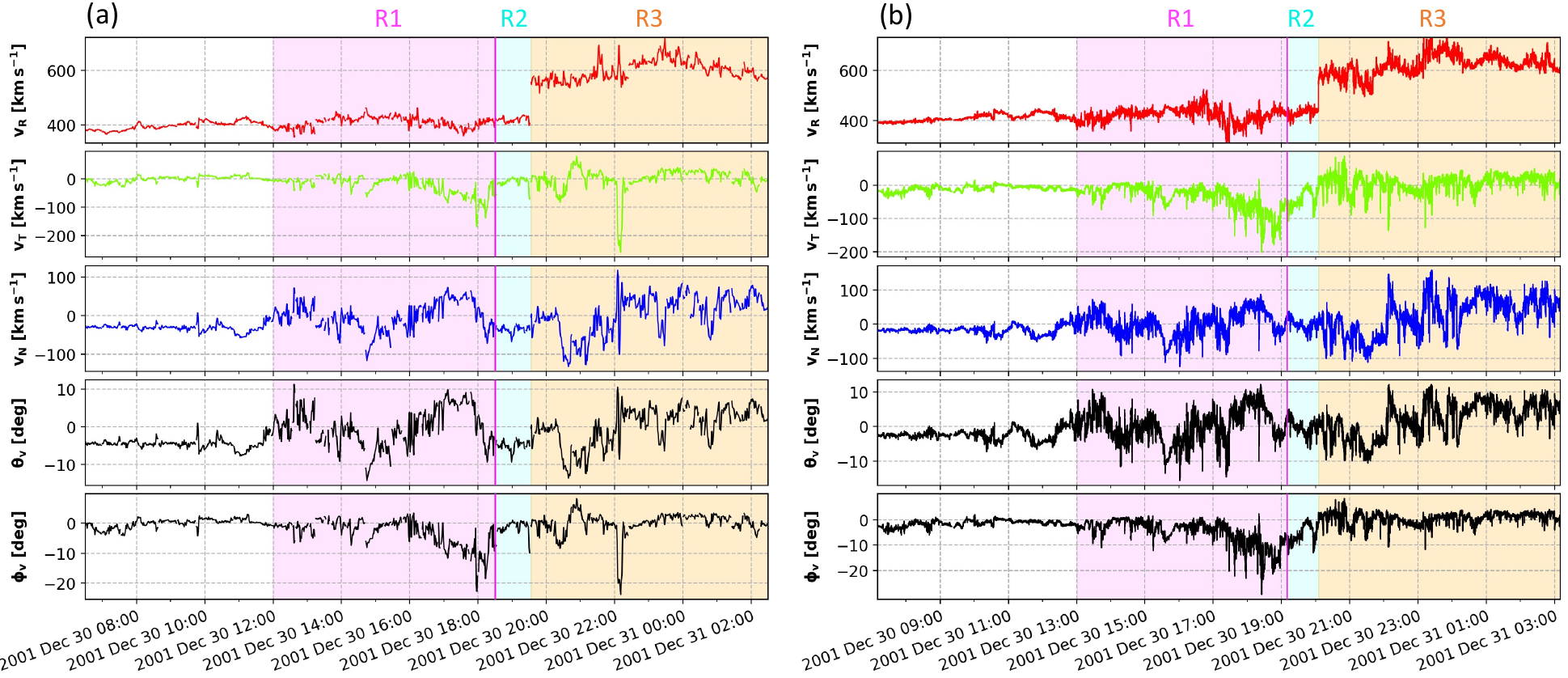}
	\caption{Velocity and non-radial flow angles in and around R1.
		(a): ACE. (b): Wind.
		From top to bottom: R, T, and N velocity components, and angle of velocity flow in the $\theta$ and $\phi$ directions.
		Regions R1, R2 and R3 are marked by the shaded magenta, cyan, and orange areas, respectively.
		The ME end time is marked by the magenta vertical line.
	}
	\label{fig:event1_velocity}
\end{figure}

The non-radial velocity signatures within the ME and following solar wind at ACE and Wind are shown in Figure~\ref{fig:event1_velocity}. Focusing on R1, Figure~\ref{fig:event1_velocity} reveals the presence of significant non-radial velocity flows in $v_T$ and $v_N$, reaching up to 100~km~s$^{-1}$ (corresponding to the north-south angles ($\theta_v$) between $\pm10^\circ$, and east-west angles ($\phi_v$) between $-20^\circ$ and $0^\circ$), which are usual in the solar wind, and even more so within MEs \citep{Al-Haddad2022}. Non-radial velocity flows above 50~km~s$^{-1}$ are observed in only about 8\% of the solar wind measurements near 1~au \citep{Al-Haddad2022}, and flows above 80~km~s$^{-1}$ only 1\% of the time. We found the velocity components in R2 are similar to those in the pre-R1 region of the ME. Non-radial velocity components are only present within R1, and given the similarity between the velocity signatures in the pre-R1 and R2 regions, it is unlikely that velocity shears are at the origin of R1. Non-radial velocity components reaching up to 100~km~s$^{-1}$ in R1 are most likely manifestations of AFs within R1, rather than causes. Moreover, if velocity shear existed, it would lead to more balanced cross helicity $\sigma_c$ \citep[e.g.,][]{soljento2023} rather than more imbalanced $\sigma_c$ in R1 as shown in Figure~\ref{fig:event1_insitu}.

% option 3: IP local formation due to the following ICME
Another possibility is that the formation of AFs in R1 is related to the presence of a following faster ICME, referred to as R3 and marked in orange in Figure~\ref{fig:event1_insitu}, which presents a forward shock and is highly Alfv\'enic.
However, it is important to note that there is a gap between the end of the ME (December 30 at 19:10 UT at Wind and 18:30 UT at ACE) and the start of the following ICME (shock times: December 30 at 20:05 UT at Wind and 19:33 UT at ACE). This implies that the following ICME did not yet have the time to directly perturb the preceding ICME at either Wind or ACE, as it had yet to reach it. This following ICME cannot therefore be responsible for the formation of R1 nor for the presence of a highly Alfv\'enic region in between the two ICMEs (indicated as R2), at least locally between ACE and Wind.
% option 3: IP remote formation due to the following ICME but somewhere else in the structure (3D effect)
It is possible, however, that the AFs in R1 were generated through the interaction with the following ICME somewhere else in the ME structure (e.g., if the following shock had already started interacting with the ME at an earlier point somewhere else in 3D space) and later propagated to Wind and ACE, marked by pin number 3 in Figure~\ref{fig:event1_source_region}~(b). Such a picture may also explain the observed AFs in the R3 region (i.e., AFs propagate from the interaction site to Wind and ACE), while R3 shares similar properties of $\sigma_c$ and magnetic components (the relatively undisturbed region after the second shock) as those measured in R1. Based on the average magnetic field strength $B$ and proton number density $n_p$ calculated throughout the ME, we estimate the average Alfv\'en speed $v_A = B/\sqrt{\mu_0 m_p n_p}$ within the ME at ACE and Wind to be around 131 and 143~ km~s$^{-1}$, respectively. The corresponding average Alfv\'{e}n travel time based on $v_A$ and on the  separation between ACE and Wind under a static assumption therefore ranges between 4.00 and 4.36~hours. This implies that in this scenario, the AFs in the R1 region must have originated more than 4~hours in the past for them to be visible at both ACE and Wind.

\section{Summary and Conclusions} 
\label{sec:conclusions}
%%%%%%%%%%%%%%%%%%%%%%%%%%%%%%%%%%%%%%%%%%%%%%%%%%%%%%%%%%%%
%{\color{blue}In some past studies, the multi-spacecraft correlations were used to quantify the coherence level of ICMEs \citep[e.g.,][]{Matsui2002,Lugaz2018,Ala-Lahti2020,Farrugia2023}, rather than using the coherence definition in \citet{Owens2017} which may not be practical for real CMEs. If combining the two definitions and AFs as shown in \citet{Scolini2023b} and this paper, ICMEs are coherent on the scale of at least the separation of the two spacecraft due to the presence of AFs, but AFs reduce the correlation of the magnetic field when the spacecraft crossing the same ME along different trajectories likely sample AFs in different oscillation phases.}

% results
We investigated an ICME arriving at 1~au on 2001 December 29, where it was detected by Wind and ACE.
The rear part of the ME (referred to as R1) exhibited large-amplitude AFs at both Wind and ACE which were associated with fluctuating non-radial flows exceeding $100$~km~s$^{-1}$.
The ME magnetic field signatures measured at ACE and Wind presented a high correlation between the two spacecraft, but when analyzing the time dependence of this correlation we found that the correlation dropped towards the back of the ME in R1.
Visually, the ME appeared characterized by a flux rope structure, but upon closer inspection the magnetic hodograms revealed a complex configuration associated with the presence of AFs. 

Our analysis revealed two key results. First, this event directly illustrates a situation where AFs decrease the correlation of the ME magnetic field profiles detected at different spacecraft at close separation, which demonstrates that AFs can make ME signatures become less self-similar along different directions by decreasing their magnetic field correlation scale. Second, we find that large-amplitude AFs can increase complexity of the ME magnetic topology detected in situ. This reveals for the first time that AFs can have direct influence on the observed magnetic topology of MEs at a given spacecraft, leading to complex magnetic field signatures and thus increasing the complexity of MEs observed in situ. We note that when the coherence as defined by \citet{Owens2017} is not practical to measure, especially for in-situ measurements,  multi-spacecraft correlations were used to quantify the coherence level of ICMEs  in past studies \citep[e.g.,][]{Matsui2002,Lugaz2018,Ala-Lahti2020,Farrugia2023}. Combining the two definitions of coherence, the investigation of AFs, and the two-spacecraft measurements as done in \citet{Scolini2023b} and this paper, ICMEs are found to be coherent on the scale of at least the separation of the two spacecraft due to the presence of AFs, but AFs reduce the correlation of the magnetic field when the spacecraft crossing the same ME along different trajectories likely sample AFs in different oscillation phases.

% AF origin
In an attempt to locate the origin of the AFs present in the rear portion of the ME, we considered a variety of observational signatures, including: the global ME configuration reconstructed from an LFF fitting of the in situ magnetic signatures; its cross helicity; the associated suprathermal electron PAD signatures; the magnetic configuration of the source region at the Sun; and the characteristics of the following solar wind. We discussed different origins including the solar origin, magnetic reconnection and velocity shear, as well as interaction with other large-scale transients. We discussed that the AFs likely formed in interplanetary space, due to interaction with a following ICME or magnetic reconnection (erosion) with the interplanetary magnetic field at the ME front. Under the first assumption, based on the spacecraft separation, we estimated that the AFs may have formed more than 4 hours in the past in order to have been visible at both ACE and Wind. 

%%%%%%%%%%%%%%%%%%%%%%%%%%%%%%%%%%%%%%%%%%%%%%%%%%%%%%%%%%%%
\begin{acknowledgments}
The views expressed are purely those of the authors and may not in any circumstances be regarded as stating an official position of the European Research Council Executive Agency and the European Commission. C.S.\ and B.Z.\ acknowledge support from the NASA ECIP program (grant no.\ 80NSSC23K1057).\
C.S.\ and N.L.\ acknowledge support from NASA grants 80NSSC20K0197 and 80NSSC20K0700.\ B.Z.\ acknowledges support from NSF AGS-2301382.
C.S.\ and R.M.W.\ acknowledge support from NASA grant 80NSSC19K0914.\
C.J.F.\ acknowledges support from NASA grant 80NSSC19K1293.\
N.M.\ acknowledges Research Foundation – Flanders (Fonds voor Wetenschappelijk Onderzoek (FWO) - Vlaanderen) for their support through Postdoctoral Fellowship 12T6521N.\
F.B.\ acknowledges support from the FED-tWIN programme (profile Prf-2020-004, project ``ENERGY'') issued by BELSPO, and from the FWO Junior Research Project G020224N granted by the Research Foundation -- Flanders (FWO).\
All data used in this study is publicly available through the NASA Solar Physics Data Facility (SPDF).\
The authors thank the ACE/SWEPAM, ACE/MAG, Wind/3DP, Wind/MAG, and SOHO/MDI instrument teams for providing the necessary data to the public.
\end{acknowledgments}





% \appendix


\bibliographystyle{aasjournal}



\end{document}